\documentclass{aa}
\usepackage{graphicx}
\usepackage{txfonts}
\usepackage{natbib}
\usepackage{ulem}

      \def\new#1 {{\bf #1 }}
      \def\cut#1 {\sout{#1} }
\begin{document}

\title{A 1.3~cm wavelength radio flare from a deeply embedded source in the Orion BN/KL region}

\author{Jan Forbrich\inst{1,2}\thanks{\email{jforbrich@cfa.harvard.edu}} \and Karl M. Menten\inst{1} \and Mark J. Reid\inst{2}}
\institute{Max-Planck-Institut f\"ur Radioastronomie, Auf dem H\"ugel, D-53121 Bonn, Germany \and Harvard-Smithsonian Center for Astrophysics, 60 Garden Street, Cambridge, MA 02138, USA}

\date{Received; accepted}

\abstract
{}
{Our aim was to measure and characterize the short-wavelength radio emission from young stellar objects (YSOs) in the Orion Nebula Cluster and the BN/KL star-forming region.} 
{We used the NRAO Very Large Array at a wavelength of $\lambda=1.3$~cm and we studied archival X-ray, infrared, and radio data.}
{During our observation, a strong outburst (flux increasing $> 10$ fold) occurred in one of the 16 sources detected at $\lambda=1.3$~cm, while the others remained (nearly) constant. This source does not have an infrared counterpart, but has subsequently been observed to flare in X-rays. Curiously, a very weak variable \textsl{double} radio source was found at other epochs near this position, one of whose components is coincident with it. A very high extinction derived from modeling the X-ray emission and the absence of an infrared counterpart both suggest that this source is very deeply embedded.}
{}


\maketitle

\section{Introduction}

A large number of compact  (sub-arcsecond size) radio sources have been found associated with stars in the Orion Nebula Cluster (ONC) and/or infrared sources in the Becklin-Neugebauer/Kleinmann-Low (BN/KL) region, an active star-forming site located behind the optically visible nebula \citep[ hereafter GMR]{gmr87} and \citep{chu87}.  Non-variable radio emission from many of the compact ONC sources comes from ionized circumstellar disks (``proplyds'') photoevaporated by the intense ultraviolet field of the ``Trapezium'' star $\theta^1$C, the exciting star of the nebula. Other sources, such as the low-mass companion to the Trapezium O-type star $\theta^1$A \citep{fel89,fel91,men07}, show pronounced variability, indicating non-thermal emission. \citet{fel93} present a comprehensive study of the Orion sources' variability on time scales from two weeks to months.

Protostellar variability has recently been extensively studied at X-ray wavelengths (most notably in the Chandra Orion Ultra-deep Project, or COUP\footnote{see the ApJS COUP Special Issue, October 2005}: \citealp{get05,fav05,fla05,wol05}), yet little is known about short-term radio variability and even less about multi-wavelength correlations. \citet{bow03} presented the first observation of a millimeter wavelength flare of a YSO in Orion (GMR~A). This observation was simultaneously covered in X-rays. The X-ray flux increased by a factor of about ten, two days before the radio flare observation of the source. During the radio flare, the X-ray flux was still at around half of the maximum level reached previously. In observations starting a few days after this outburst, the source was found to show millimeter wavelength activity for at least 13 days, while the infrared $K$-band magnitude remained unchanged \citep{fur03}.

While longer-term radio variability of YSOs has been known for some time (e.g. \citealp{fem99}), there are few radio observations of a YSO that cover a period of strong variability, i.e., that document an outburst. Here, we present a new event of this kind occurring in the BN/KL region. In an 8 hour VLA observation at a wavelength of 1.3~cm, we serendipitously found extremely strong variability in a source that had shown only very weak, if any, emission at other occasions and/or at other wavelengths. While this source is detected in X-rays it was not detected at infrared wavelengths. This implies that is is deeply embedded, contrary to GMR~A, which is a bright near-infrared source \citep{bow03}.

The special observing mode employed and the data reduction are described in Sect.~\ref{obs}. The outburst is described in Sect.~\ref{results}, which also presents observations of this source at other wavelengths and discusses the physical scenario. We give a summary in Sect.~\ref{summary}.

\section{\label{obs}Observations}
\subsection{Radio data: VLA observations}

In the 1990s several VLA observations of the Orion region were made at X band (8.4~GHz) and K band  (22.285~GHz). The K-band observations discussed here were conducted on 1991 July 5 with the NRAO \footnote{The National Radio Astronomy Observatory (NRAO) is operated by Associated Universities, Inc., under
a cooperative agreement with the National Science Foundation.} Very Large Array in its highest-resolution A configuration. On-source observation time started at 13:23 IAT (01:05 LST) and lasted until 22:02 IAT (09:45 LST). Observations started with clear skies, a temperature of 12.7$^\circ$C and winds of 0.2~m/s from noth-west. Eight hours later, at 21:36 IAT, the temperature was at 29.3$^\circ$C, with winds at 3.52~m/s from the east. By then, the sky was 35~\% cloudy, which was also the case after the end of the observation, at 23:17 IAT, when the temperature was at 28.9$^\circ$C with winds of 4.6~m/s from the southeast. During the entire time, barometric pressure fell from 794.4~mbar to 791.3~mbar. We discuss two additional VLA observations carried out in X-band on 1991 September 6 (8.5 hours duration), as well as on 1994 April 29 (10 hours duration). In both cases, the VLA also was in its A configuration and the standard continuum setup with a bandwidth of $2\times50$~MHz (RCP+LCP each) was used. For details on the analysis of the 1994 dataset, see \citet{mer95}. A very similar method was used for the 1991 X-band dataset.

\subsubsection{Basic calibration and imaging}
The Orion YSO sources have weak radio emission, with typical flux densities $\sim$5 mJy, and ``conventional'' self-calibration is impossible at K band. Instead, we make use of bright, nearby water maser emission (e.g., \citealp{gau98}). We calibrated the data in the same way first described by \citet[ see also the Appendix of \citealp{rem97} for a detailed description]{rem90}. In short, a dual intermediate-frequency (IF) band setup was used with a narrow IF band (3.125 MHz) covering the $6_{16}\to5_{23}$ H$_2$O maser line (rest frequency 22238.08 MHz) and a broad IF band (50 MHz) centered at a 50 MHz higher frequency on a line-free portion of the spectrum. A position near the infrared source IRc~2 was observed in both bands simultaneously. Observations of the quasars 0501--019 and 0535+135 were interspersed alternately every $\sim 30$ minutes to determine electronic phase offsets between the bands. The phase center position was $(\alpha,\delta)_{\rm J2000}$ = $05^{\rm h}35^{\rm m}14{\rlap.}^{\rm s}479, -05^{\circ}22'30{\rlap.}$\,$''$\,$\!568$.

Absolute flux calibration was obtained from observations of 3C286 using flux densities interpolated from the values given by \citet{baa77}. The \textit{uv}-data were edited, removing obviously faulty visibilities.

The narrow-band data were then ``self-calibrated'' with the very strong maser signal as a phase reference. The phase and amplitude corrections were then applied to the broadband data, and a high-quality map of the continuum emission was produced. Once the continuum data were ``cross-self-calibrated'' the data were imaged in the usual way with NRAO's Astronomical Image Processing System (AIPS), setting the IMAGR mapping task parameter ROBUST to zero. The half-power beam-width of the synthesized beam was $109 \times 97$ milliarcseconds (mas) at a position angle of $37^\circ$ (E of N). The map size was $8192 \times 8192$ times 20 mas pixels, which covers an area of $164\times164$ arcseconds, covering most of the primary beam of a single VLA antenna.

\subsubsection{Further Processing}
We examined and identified a number of compact sources in the continuum image. However, there was an uneven distribution of noise over parts of the image. Checking our calibration and editing, we found no apparent error in our processing. We then produced two images, one from the first (temporal) half of the \textit{uv} data set and one from the second. Comparing the sources' flux densities in each of the images, we noted that one source had different flux densities in these maps, suggesting significant variability. We call this source RBS~J053514.67-052211.2 for ``radio burst source'', in the following referred to as the RBS. Its variability caused the unusual noise.
Imaging the entire \textit{uv} data, the deepest continuum map has an rms noise level of 0.18~mJy. The AIPS task SAD was used to find all sources exceeding six times the noise level in that map and fit Gaussian functions to detected sources. The 16 detections were then studied in more detail using the AIPS task JMFIT, which yielded estimates of flux density, size, and position. Flux densities were then corrected for primary-beam attenuation. From a comparison with VLBI detections \citep{men07}, we estimate the absolute position accuracy to be about $0\farcs1$.

In order to be able to study the surroundings of the RBS in more detail by attempting to remove the ``noise'' induced by the variable source, we then also produced a RBS-subtracted map. We split the cross-self-calibrated data into one-hour long chunks, which were mapped individually. Then, UVMOD was used to subtract a point source model with the position and flux density determined for the RBS in each of the hour-long segments of data before using DBCON to concatenate the resulting \textit{uv} data subsets together.

\begin{table*}
\begin{center}
\caption[]{Sources detected in the K-band dataset.}
\begin{tabular}{llrrrrrll}
\hline
& Position & $r^{\rm a}$ & Peak brightness & int./peak & $5\sigma$ p$_{\rm C}$ & ID$^{\rm b}$ & COUP$^{\rm c}$ & $\Delta\Theta[$''$]^{\rm d}$ \\
& ($\lambda=1.3$~cm, J2000.0)  & ['] & [mJy/bm] & ratio & [\%]  &    &      &  \\
\hline
\hline
 1 & 5 35 11.801 --5 21 49.15 & 0.96 & $  8.45 \pm 0.37 $ & $1.34 \pm 0.15$ & $<24$ & GMR A  & 450 & 0.08 \\
 2 & 5 35 14.116 --5 22 22.81 & 0.16 & $ 11.59 \pm 0.19 $ & $1.43 \pm 0.06$ & $< 9$ & BN   & --  & --	\\
 3 & 5 35 14.507 --5 22 30.40 & 0.01 & $  3.1  \pm 0.18 $ & $1.84 \pm 0.26$ & $<32$ & GMR I$^{\rm e}$  & --  & --   \\
 4 & 5 35 14.665 --5 22 11.18 & 0.33 &     --$^{\rm f}$   &   --$^{\rm f}$  &	--$^{\rm f}$   & RBS& 647 & 0.09 \\
 5 & 5 35 14.895 --5 22 25.30 & 0.14 & $  1.53 \pm 0.19 $ & $1.00 \pm 0.34$ & $<65$ & GMR D  & 662 & 0.10 \\
 6 & 5 35 15.769 --5 23 09.81 & 0.73 & $  1.64 \pm 0.27 $ & $1.36 \pm 0.58$ & $<93$ & GMR 25 & 732 & 0.14 \\
 7 & 5 35 15.820 --5 23 14.01 & 0.80 & $ 11.92 \pm 0.30 $ & $1.16 \pm 0.08$ & $<13$ & GMR 12 & 745 & 0.23 \\
 8 & 5 35 15.836 --5 23 22.38 & 0.93 & $  3.07 \pm 0.33 $ & $3.57 \pm 0.87$ & $<65$ & GMR 11 & 746 & 0.16 \\
 9 & 5 35 16.065 --5 23 24.20 & 0.98 & $  2.48 \pm 0.36 $ & $2.86 \pm 0.97$ & $<86$ & GMR 8	& --  & --     \\
10 & 5 35 16.070 --5 23 06.97 & 0.72 & $  1.01 \pm 0.26 $ & $3.48 \pm 1.97$ & $<100$ & GMR 15 & 766 & 0.19 \\
11 & 5 35 16.288 --5 23 16.49 & 0.89 & $  3.26 \pm 0.32 $ & $3.10 \pm 0.69$ & $<55$ & GMR 7  & 787 & 0.06 \\
12 & 5 35 16.326 --5 23 22.53 & 0.98 & $  2.88 \pm 0.38 $ & $1.58 \pm 0.52$ & $<72$ & GMR 16 & --  & --   \\
13 & 5 35 16.398 --5 22 35.23 & 0.48 & $  3.09 \pm 0.22 $ & $0.97 \pm 0.19$ & $<38$ & GMR K  & --  & --   \\
14 & 5 35 16.750 --5 23 16.36 & 0.95 & $  4.50 \pm 0.34 $ & $5.17 \pm 0.85$ & $<46$ & GMR 6  & 826 & 0.11 \\
15 & 5 35 16.846 --5 23 26.12 & 1.10 & $  3.15 \pm 0.44 $ & $4.71 \pm 1.45$ & $<84$ & GMR 5  & --  & --   \\
16 & 5 35 18.367 --5 22 37.36 & 0.97 & $ 13.95 \pm 0.39 $ & $1.19 \pm 0.09$ & $<14$ & GMR F  & 965 & 0.17 \\
\hline
\label{dbconnedlist}
\end{tabular}
\end{center}
$^{\rm a}$ distance from phase center, in arcminutes. One arcminute is approximately the radius of the half-power beam width at 22~GHz. \\
$^{\rm b}$ The numbers/letters Orion radio source nomenclature was introduced by \citet{gmr87} who used numbers to designate 14 sources around the Trapezium cluster and letters A--G for sources associated with the Orion molecular cloud. These lists were extended by \citet{gar87,gar89}, \citet{fel93} and \citet{zap04a}. Note that despite its designation, the optically visible GMR~A is most likely a member of the ONC.\\
$^{\rm c}$ Counterpart number in the data of the Chandra Orion Ultra-deep Project.\\
$^{\rm d}$ Angular separation of X-ray source and $\lambda=1.3$~cm radio source in arcseconds (if $<0.3''$, used as coincidence limit).\\
$^{\rm e}$ See \citet{mer95} and \citet{rei07} for an in-depth discussion of source I.\\
$^{\rm f}$ The position of RBS~J053514.67-052211.2 was determined separately at the flare maximum. The source was unresolved at all times. The upper limit for the $\lambda=1.3$~cm flux density of the north-eastern component of the double X-band source is 1.14~mJy (3$\sigma$), see text.\\
\end{table*}

\section{Results}
\label{results}

\begin{figure} 
\begin{center}
\includegraphics*[width=\linewidth,bb=40 140 550 670]{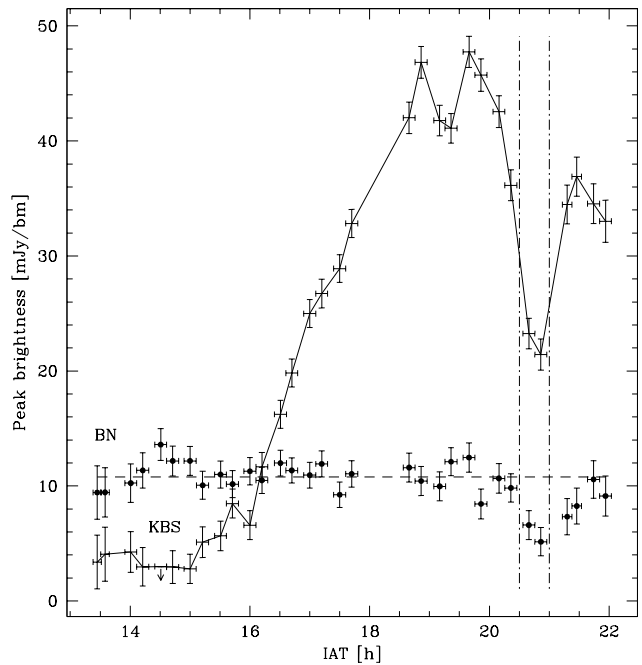}
\caption{ Light curves of the flaring RBS (solid line) and BN (circles). Data were taken on 1991 July 5. For the determination
of the error bars as well as for information on the suspect data
between the two vertical dash-dotted lines, see text. The data for the RBS
contain one $3\sigma$ upper limit, denoted by an arrow, and the dashed
line indicates the weighted mean of the BN peak brightness values. } 
\label{JMFITHR}
\end{center}
\end{figure}

\begin{figure}[h!]
   \centering
   \includegraphics*[width=7.6cm,bb=60 201.36 550 670]{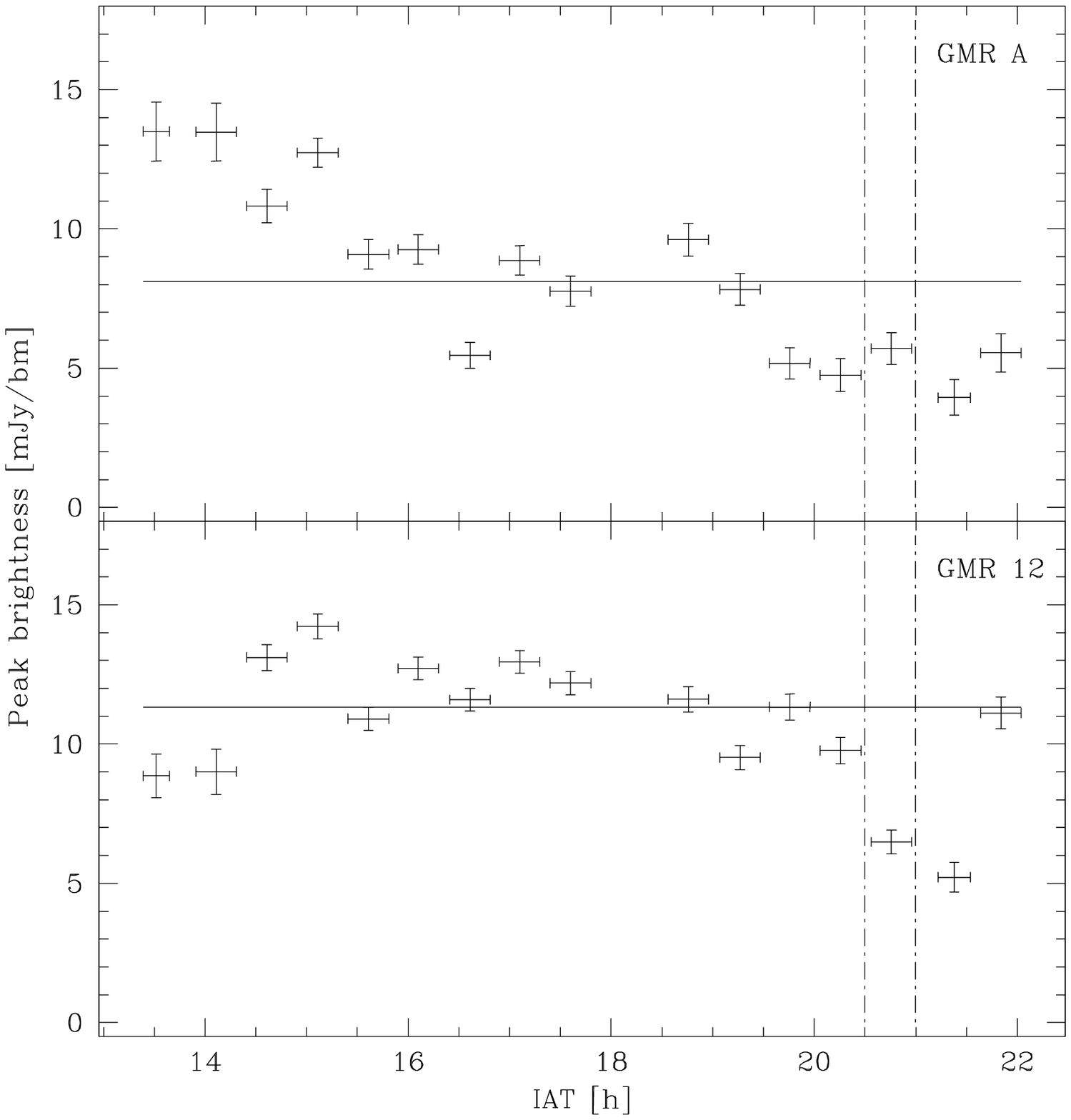}
   \includegraphics*[width=7.6cm,bb=60 201.36 550 660.02]{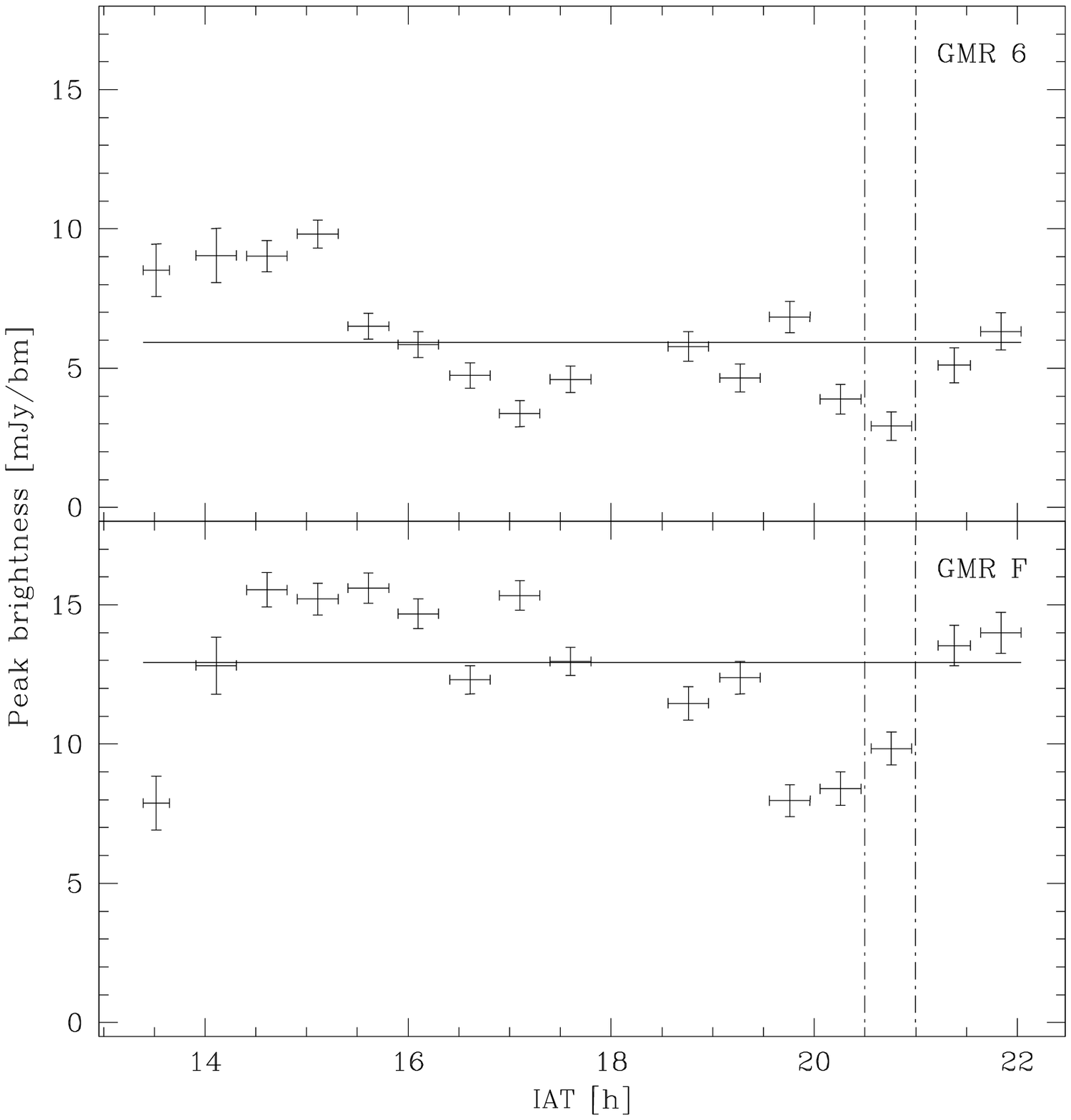}
   \includegraphics*[width=7.6cm,bb=60 396 550 660.02]{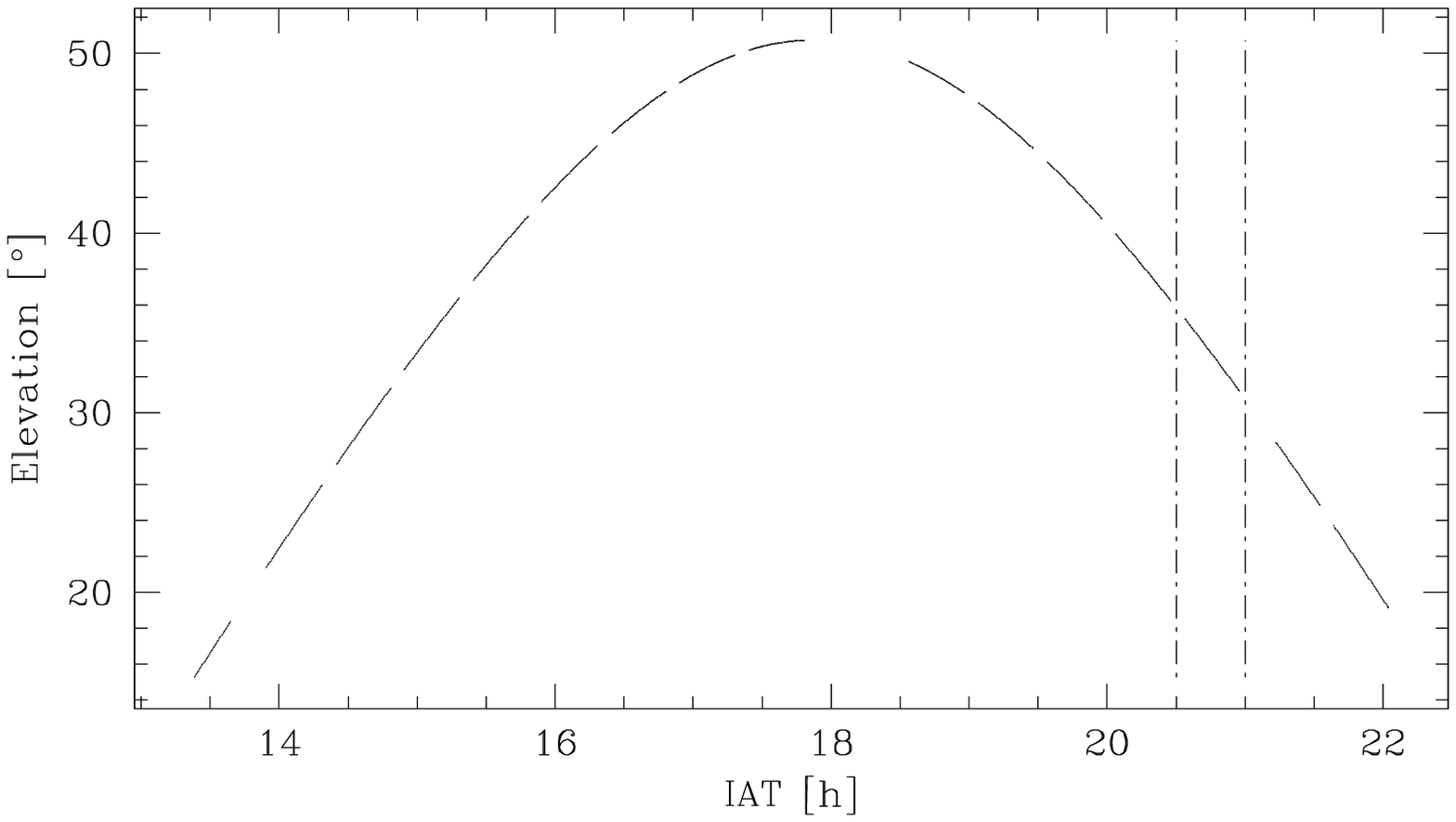}

\begin{center}
\caption{ 1.3 cm wavelength flux densities from our observation on 1991 July 5 for the brightest sources (as labeled), except for the RBS and BN. The lowest panel shows the source elevation. Horizontal error bars mark the time ranges. Indicated for each source is the weighted mean of the measurements calculated without the one data point per light curve marked suspect. For the determination of the error bars as well as for information on the suspect data between the two vertical dashdotted lines, see text. }
\label{JMFITQ}
\end{center}
\end{figure}

Out of the detected sources, only the burst source RBS had not previously been identified at radio wavelengths. No circular polarization was found in any source. For the RBS, the 5$\sigma$ upper limit drops to $p_C <5.9$~\% during the flare maximum.
The results of the source identification are shown in Table~\ref{dbconnedlist}. Listed are the 16 source positions with the respective distances from the phase center, spatial peak brightness (corrected for primary beam attenuation), ratios of integrated flux density to peak brightness, 5$\sigma$ upper limits for circular polarization as well as previous source designations. The median offset from the respective positions given in \citet{zap04a} is $0\farcs11$. The quoted flux density errors are those calculated by JMFIT, i.e., consisting basically of the local map noise scaled by the correction for primary beam attenuation. The position of the RBS was determined separately at the flare maximum.

Variability on timescales of hours can be expected from non-thermal radio sources. While circular polarization was not detected here and thus cannot be used to discriminate between thermal and non-thermal emission, it is worth looking at the ratios of integrated flux density to peak brightness for all sources as a secondary criterion: Non-thermal sources are unresolved at the angular scales considered here. The fact that GMR~A, GMR~12, and GMR~F were detected by VLBI \citep{men07} and thus have a significant non-thermal component suggests that sources with ratios of below about 1.3 probably are non-thermal (Table~\ref{dbconnedlist}). Indeed, these are the three sources for which \citet{fel93} report highly variable radio spectral indices. With a variability factor of 43, GMR~A is the most variable source in the multi-year X-band dataset of \citet{zap04a}, and also GMR~12 and GMR~F are variable, if much less (by factors of 4 and 2, respectively). Six sources were bright enough for a meaningful study of time variability: RBS, BN, GMR~A, GMR~12, GMR~6, and GMR~F. For this analysis, we mapped the entire dataset in different time intervals, starting with chunks having durations of about one hour. The RBS turned out to be bright enough for imaging in intervals of about 15~min at noise levels of typically 1~mJy (Fig.~\ref{JMFITHR}). GMR~A, GMR~12, GMR~6, and GMR~F were imaged at about 30~min time resolution (Fig.~\ref{JMFITQ}).

We checked for elevation-dependent systematic errors in our time sequence flux density measurements, since the source elevations varied between $15^\circ$ and $50^\circ$ (see Fig.~\ref{JMFITQ}). BN is known to be a thermal source and thus it should be constant over the short time periods we are considering. As can be seen in Fig.~\ref{JMFITHR}, BN displays no obvious trends that would correlate with source elevation nor any strong indication of variability. It is also only $0\farcm16$ off the phase center. We thus fitted a constant emission level to the light curve of BN, using the errors determined by JMFIT from the map noise. The suspect data discussed below was not taken into account. In order to estimate realistic errors, we increased the size of the error bars so that the reduced $\chi^2$ would be 1, leaving out clearly suspect data (see below). This is accomplished with errors 1.47 times larger than the initial value. The reasons for this correction are not exactly known. Pointing errors could play a role at this high observing frequency. Subsequently, this factor was also applied to the error bars in the light curves of the other sources.

While the RBS is clearly variable beyond doubt, the other sources appear to be nearly constant. The sources shown in Fig.~\ref{JMFITQ} are all relatively far off the phase center (see Table~\ref{dbconnedlist}) and thus vulnerable to pointing errors, complicating the analysis of their variability. Since GMR~A, GMR~12, and GMR~F have all been shown to emit non-thermal radio emission by respective VLBI detections, at least some of their variability may be real. Surprisingly, also GMR~6 appears to be marginally variable. This source has been described as thermal source before \citep{fel93,zap04a} and is clearly resolved in our data (Table~\ref{dbconnedlist}). The fact that GMR~6 still appears to be variable very likely is predominantly due to the fact that it was observed close to the radius of the half-power beamwidth where small pointing errors would result in large flux density variations.

\begin{figure*}
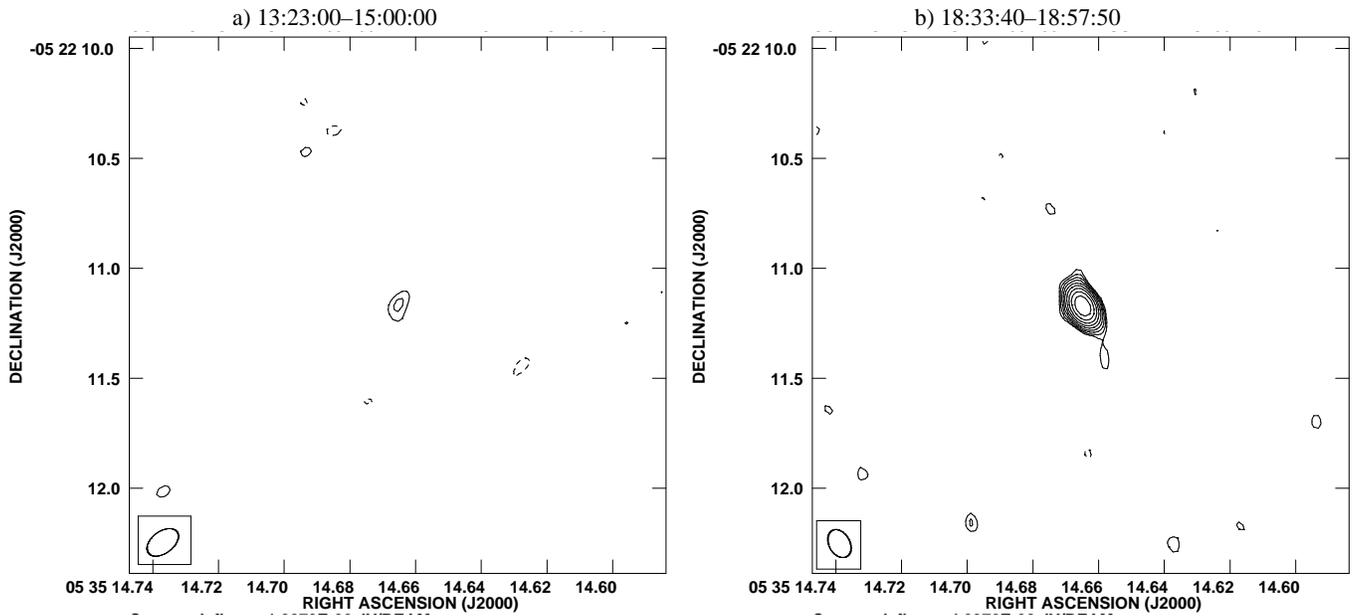

\begin{minipage}{9cm}
  \centerline{a) 13:23:00--15:00:00}
  \includegraphics*[width=9cm,bb=40 181 570 637]{8070fig3a.ps}
\end{minipage}
\begin{minipage}{9cm}
  \centerline{b) 18:33:40--18:57:50}
  \includegraphics*[width=9cm,bb=40 181 570 637]{8070fig3b.ps}
\end{minipage}

\caption{Maps of the flare source, using (a) all the data up to 15:00:00 IAT, when the source is weakest, and (b) the data form 1991 July 5, 18:33:40 to 18:57:50 IAT containing the light curve maximum in Fig.~\ref{JMFITQ}. The contours denote  $-2,-\sqrt{2}, -1, 1, \sqrt{2}, 2, ...$ times the map $3\sigma$ rms noise levels of 1.52~mJy and 1.79~mJy, respectively.}
\label{flaresrcmaps}
\end{figure*}

The RBS is detected as a weak source at the beginning of the observation. Maps of the source and its surroundings are shown in Fig.~\ref{flaresrcmaps}. Following the steep rise of the flare, a maximum of 47~mJy is reached about five hours after the beginning of the observation.  Subsequently, the flux density varies a bit before falling to 20~mJy at 20:30 IAT before rising again to nearly 40~mJy. What caused the drop around 20:30 IAT remains unclear. The source elevation at that time was reasonably high at $\sim35^\circ$. However, since BN in particular, but also some other nearby sources experienced a similar drop at that time,  this suggests an instrumental or observational effect. In Figs.~\ref{JMFITHR} and \ref{JMFITQ}, we marked data as suspect because BN deviated from the median value by more than $3\sigma$. At all times, the RBS is unresolved. The steep rise by a factor of 2 in about 15 minutes implies that the emitting region is smaller than $\approx15$ light-minutes, i.e. $<2$~AU.

\subsection{Additional radio data: VLA 8.4~GHz archival data}

While the discovery observation on 1991 July 5 was not accompanied by observations at other frequencies, the Orion region has been frequently re-observed since then. One subsequent observation was carried out on 1991 September 6 in X-band (8.4~GHz). Analyzing the data, we discovered  a \textsl{double} radio source with a separation of $0\farcs45$ at the position of the RBS; see Table~\ref{xbandcoord} for the coordinates. The RBS's position determined from our 22~GHz data is within 20~mas of the SW component of the double structure seen at X-band. No 22~GHz emission was found toward the north-eastern X-band source. A reasonable upper limit to its $\lambda=1.3$~cm flux density is the 3$\sigma$ rms noise level at its position in the combined, RBS-subtracted map which is 1.14~mJy in a box with a side length of $0\farcs4$ (influenced by the residuals of the RBS removal).

\begin{figure}
\includegraphics*[width=\linewidth, bb= 20 43 540 552]{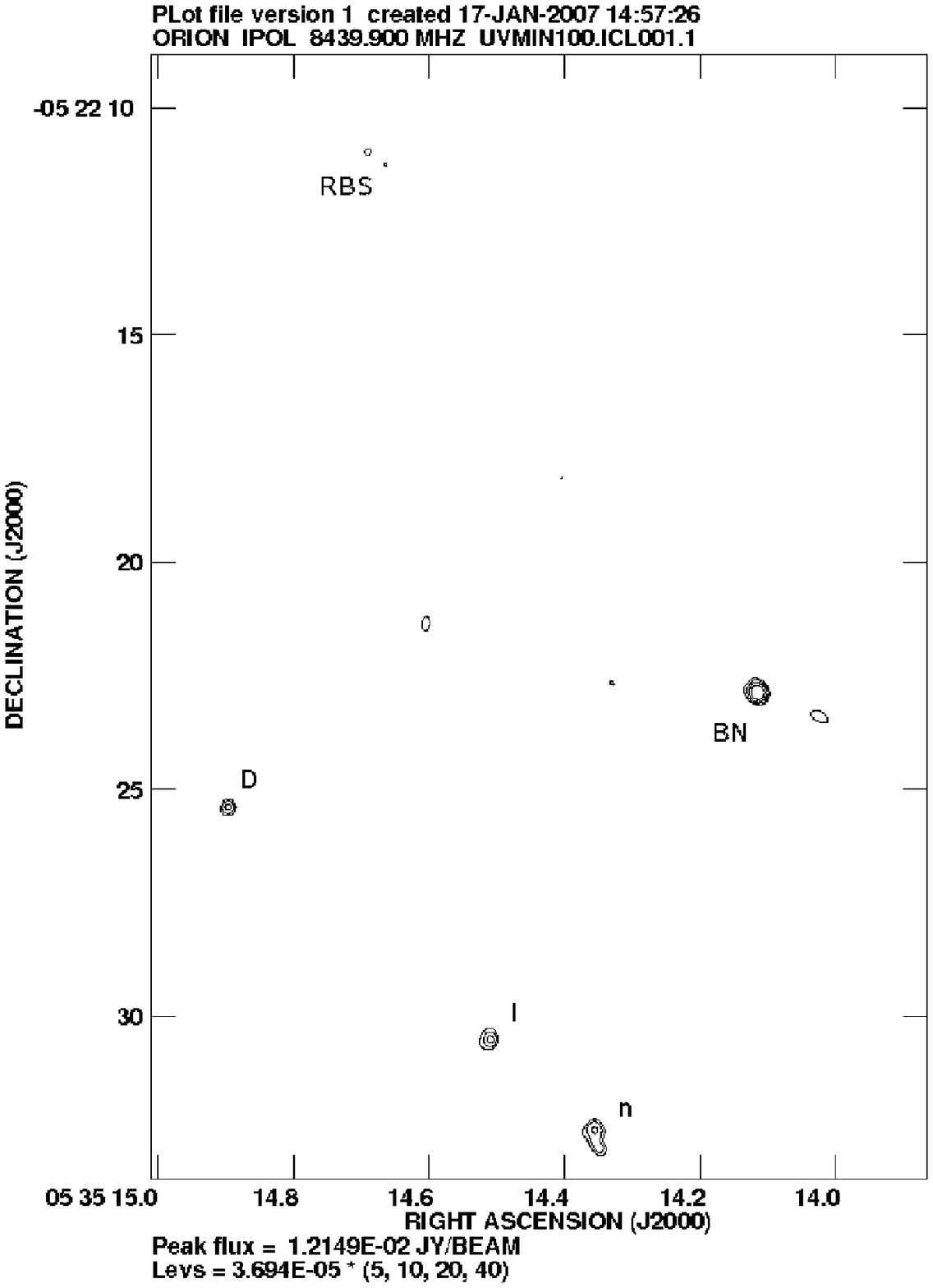}
\caption{8.4~GHz image of the region as observed on 1994 April 29 \citep{mer95}. Contour lines are 5, 10, 20, and 40$\sigma$ with an rms noise of $\sigma=37~\mu$Jy (see \citet{mer95} for a more detailed analysis of this dataset). The double X-band source is seen in the upper left corner. The RBS is associated with the south-west component; the bright source right of the center is BN. \label{xbandplot}}
\end{figure}

While a full discussion of all subsequent observations is beyond the scope of this paper, we note that in X-band (8.4~GHz) observations carried out on 1994 April 29, \citet[ see their Fig.~1 in the upper left corner]{mer95} detect this double source again (Fig.~\ref{xbandplot}). Interestingly, while all other sources discussed here and listed in Table~\ref{dbconnedlist} have X-band radio counterparts in multi-epoch data ranging from 1994--1997 discussed by \citet{zap04a}, this double X-band structure was not reported. This may be due to the fact that the flux densities in the two epochs during which the double source was detected in X-band are close to the significance threshold and may be below it depending on the noise level achieved when processing the data.

\begin{table}
\begin{center}
\caption[]{The double 8.4~GHz counterpart to the RBS as observed on 1991 September 6 and on 1994 April 29.}
\begin{tabular}{llll}
\hline
Component & Position (J2000.0) & \multicolumn{2}{l}{Flux density$^{\rm a}$} \\
          &                    & 1991 & 1994\\
\hline
\hline
SW & 05 35 14.664 --05 22 11.16 & 0.28& 0.22\\
NE & 05 35 14.688 --05 22 10.89 & 0.23& 0.32\\
\hline
\label{xbandcoord}

\end{tabular}
\end{center}
\vspace{-5mm}
$^{\rm a}$ peak brightness in mJy/bm
\end{table}

\subsection{X-ray data: Archival Chandra observations}

\begin{figure}
 \includegraphics*[width=\linewidth]{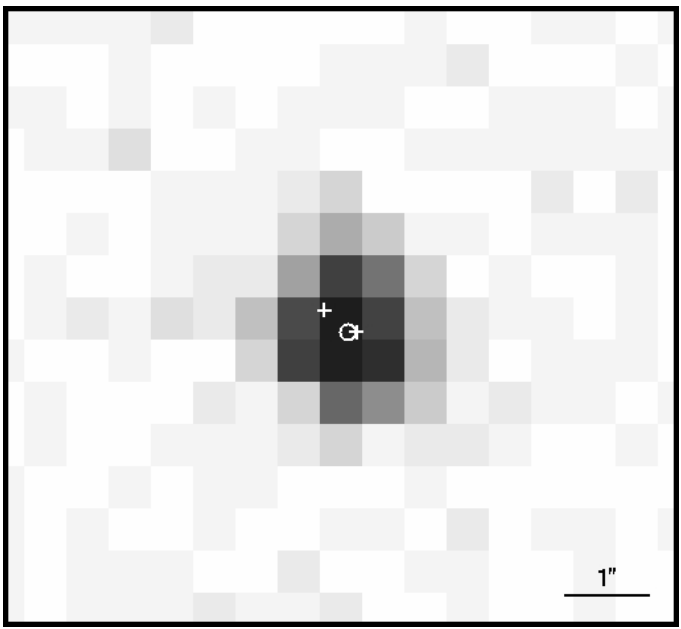}
\caption{COUP X-ray image of the RBS. The peak positions of the two 3.5~cm radio sources are shown as crosses and the 1.3~cm wavelength flare source as a circle. Symbols are larger than the positional uncertainties.\label{COUP647plot}}
\end{figure}

The Orion Nebula Cluster and KL region have been extensively studied at X-ray wavelengths, most notably in the Chandra Orion Ultradeep Project, an 838~ksec observation carried out over a period of 13 days in 2003 January.  The RBS has an X-ray counterpart which has been catalogued as X-ray source COUP~647 or CXO ONC J053514.6-052211. Fig.~\ref{COUP647plot} shows an image of the source from the COUP data, together with the positions of the two 8.4~GHz sources and the RBS flare source. The radio flare source seems to be coincident with the X-ray flare source to within $0\farcs10$,  i.e., within the mutual uncertainties. The COUP light curve shows a weak X-ray source with a quiescent flux corresponding to count rates of around 1~ksec$^{-1}$ and a considerable flare reaching 17~ksec$^{-1}$ shortly before the end of the observation \citep{get05}. The observation ends with the count rate still at about half the maximum level, the flaring state extending for less than 10~ks at
the end of the observation.

With the 808 total X-ray counts from this source the absorbing column density was determined to be log(N$_{\rm H}$ [cm$^2$]) = 23.51. Using the empirical relation N$_{\rm H}$ [cm$^2$]$\approx 2\times 10^{21}\times A_V$~[mag] \citep{ryt96,vuo03}, this corresponds to a very high visual extinction, $A_V$, of $\approx 160$~mag,  showing that this is a deeply embedded source and ruling out any observable infrared counterpart. The X-ray emission is quite hard with a derived plasma temperature of $kT=5.5$~keV or 64~MK. The total X-ray luminosity ($0.5-8$~keV) corrected for absorption is 10$^{30.9}$ erg s$^{-1}$. \citet{tsu05} find that COUP 647 is one out of only seven
sources in the entire COUP sample with detectable fluorescent 6.4~keV iron line emission and among these the only source without a near-infrared counterpart.  This emission is interpreted as reprocessed radiation from an X-ray illuminated circumstellar accretion disk.

\citet{fei02} analyze two older \textsl{Chandra} ACIS-I datasets from 1999 and 2000, covering 86~ksec in total. They detect a weak source (ID 428) with a total of 74 counts and classify it as long-term variable due to differences between the two observations (see also \citealp{gar00} for the first of the two datasets).
\citet{fla03} detect the source in \textsl{Chandra} HRC observations in 2000 February; they find 15 counts in a 63~ksec exposure. Thus, in a total of 987~ksec (i.e. 11.4 days) of \textsl{Chandra} X-ray observations, only a single flare was observed (unfortunately, the source falls near the limits of the observed fields in archival XMM-\textsl{Newton} data).

\subsection{Infrared data: Archival HST, VLT, and \textsl{Spitzer} data}
\label{oriir}
Of the 16 sources we detected, twelve have near-infrared counterparts; only the RBS and GMR I, D, and K do not. Two of these sources (GMR I and D) are close to BN. GMR~I is the radio source associated with the infrared structure IRc~2, which may be one of the dominant energy sources in the region (but see \citealp{mer95}).

The RBS remains undetected in near-infrared $JHK$ data taken with the ESO Very Large Telescope (as quoted in the COUP Atlas with 5$\sigma$ point source limiting magnitudes of approximately 22, 21, and 20 at $J_s$, $H$, and $K_s$, \citealp{get05}). This is certainly due to the high absorbing column density causing an extinction of $A_V \approx 160$~mag (see above). The source remains undetected as well in HST-NICMOS observations\footnote{http://nicmosis.as.arizona.edu:8000/TRAPESIUM/ TRAPEZIUM\_DOWNLOAD.html} performed at wavelengths of 1.1~$\mu$m and 1.6~$\mu$m, reaching a limiting magnitude of $\sim 17$~mag in the
latter band. No detection was reported by \citet{rob05} from their mid-infrared observations at 10~$\mu$m and 20~$\mu$m. As an additional check, we downloaded archival \textsl{Spitzer}-IRAC mid-infrared observations (all four bands). No source is detected at the position of the flare source (however, the sensitivity of that data is limited by the very bright neighboring BN object).

\subsection{Emission mechanism}

Absent further information on this flare, it is difficult to elaborate on the physical mechanism causing it. The key information that we have from our observation concerns the timescale of variability, the flare rise time of several hours, and the fact the observed radiation is not significantly circularly polarized. As noted above, the flare rise time indicates an emission region smaller than 2~AU, while the constraint from the fact that the source remains unresolved in the VLA data yields a source size of $<0\farcs1$, i.e. $<41$~AU (using $d=414$~pc, \citealp{men07}). Assuming a source radius of $\approx 1$~AU, the maximum flux density of the RBS corresponds to a brightness temperature of 7.6~MK, indicating non-thermal emission \citep{gue02}. Unfortunately, we do not have simultaneously obtained multi-band data to assess the spectral index. The role of the apparently variable double 3.5~cm radio source remains unclear. At 190~AU, its separation is larger than the scales discussed previously.

\section{Summary}
\label{summary}

We report the observation of a 1.3~cm wavelength radio flare of a deeply embedded source in the 
Orion BN-KL region. The flare source was observed to rise within four hours to become the strongest source in the field. This was followed by a decrease in flux density.

The source does not have a near-infrared counterpart in VLT and HST data but does have a hard X-ray counterpart (a flaring COUP source), implying an absorbing column density corresponding to $A_V \approx 160$~mag. Consistent with this $A_V$, the source remains undetected at mid-IR wavelengths, although deep observations of the region are hampered by the bright neighboring BN object.

There are 15 additional sources detected at $\lambda=1.3$~cm, including GMR~A, towards which a 86~GHz flare was reported by \citet{bow03}. Apart from BN and the RBS, variability was studied for the bright sources GMR~A, GMR~12, GMR~6, and GMR~F. These sources were found to be marginally variable, but instrumental effects appear to play a role. All 16 sources detected at $\lambda=1.3$~cm have counterparts at $\lambda=3.5$~cm, including the flare source whose 3.5~cm wavelength counterpart matches one component of a double structure with a separation of  $0\farcs45$. The very weak components of the double are variable and have been detected on two different observations, but undetected at other times.

The RBS is the second YSO in Orion known to show burst behavior at radio as well as at X-ray wavelengths. It is much deeper embedded than GMR~A \citep{bow03} and thus may be in an earlier evolutionary stage. In our data, GMR~A is detected at a flux density level of about 8~mJy, roughly consistent with the post-flare flux density observed by \citet{bow03}. While for the RBS, two separate flares have been observed in X-rays and at radio wavelengths, both observations ended before a quiescent level was reached. During the COUP observations, GMR~A showed a more complex lightcurve, but we may see only the beginning of an activity phase in the COUP data of the RBS.

\begin{acknowledgements}
We would like to thank an anonymous referee
for helpful and constructive comments that lead to improvements of
this paper. We would also like to acknowledge Thomas Preibisch and Andreas Brunthaler for advice and discussions.
\end{acknowledgements}

\bibliographystyle{aa}
\bibliography{bibmasterKMM}

\end{document}